\documentclass[reprint,amsmath,amssymb,nofootinbib]{revtex4-2}

\usepackage{graphicx}
\usepackage{color}
\usepackage[dvipsnames]{xcolor}
\usepackage{amssymb}
\usepackage{amsfonts}
\usepackage{xcolor}
\usepackage{amsmath}
\usepackage[utf8]{inputenc}

\usepackage{hyperref}
\usepackage{float}
\usepackage[normalem]{ulem}
\usepackage{siunitx}

\hypersetup{colorlinks = true, linkcolor=blue, citecolor=orange, urlcolor=blue, bookmarksnumbered = true}

\begin{document}

\title{Detailed Analysis of the Superconducting Gap with Dynes Pair-Breaking Scattering\\\vspace{0.2cm}
}

\author{Anastasiya Lebedeva}
\author{František Herman}
\email{herman2@uniba.sk}
\affiliation{Department of Experimental Physics, Comenius University, Mlynská Dolina F2, 842 48 Bratislava, Slovakia}


\begin{abstract}
We study the energy gap within the Dynes superconductor theory. This model generalizes the Bardeen-Cooper-Schrieffer (BCS) approach by including the pair-breaking scattering, introducing the tunneling in-gap states up to a Fermi level. We analytically solve the energy gap equation in various limit cases. The solution provides simple tools for further studies, compared to more complex numerics, and highlights the basic characteristics of the theory. First, in the critical limit of pair-breaking scattering, we derive an analytical form of zero-temperature gap to transition temperature ratio. Next, we derive the dependence of the energy gap close to critical temperature and look at its behavior for general and critical pair-breaking scattering rate. Furthermore, we compare our result with the numerical solution of the gap equation assuming general temperature. We show the range of temperatures, for which the analytical approximation is valid. In the end, we provide the approximative formula of the gap, assuming general pair-breaking scattering, emphasizing its exact behavior close to transition temperature.
\end{abstract}

\maketitle

\section{Introduction}
\label{intro}
The Dynes superconductor theory in general combines the effects of the superconducting electron pairing together with the pair-breaking and pair-conserving scattering in a simple fashion~\cite{Herman16, Herman17_1, Herman17_2, Herman18}. Its formulation based on Green's functions solving the coherent potential approximation (CPA) equations~\cite{Soven67, Velicky68, Weinkauf75} within the superconducting state is convenient for the experiment from several points of view. Let us mention, e.g., the tunneling measurements targeting the superconductor-insulator transition~\cite{Szabo16, Zemlicka20}, next the analysis of the superconducting specific heat jump in $\mathrm{La_2Ni_2In}$~\cite{Mainwald20}, or optical conductivity study in $\mathrm{NbN}$ focusing on the response to external electromagnetic field~\cite{Sindler22}. I.a. the framework allows estimating the microscopic properties, like the fundamental scale of the size of a Cooper pair~\cite{Herman23}.

Originally, the authors in Ref.~\cite{Herman16} used the Lorentzian form of pair-breaking scattering probability distribution $P_{\Gamma}(E)$ with the width $\Gamma$\footnote{Using units $\hbar=k_B=1$.} around the Fermi energy $E=0$. This step was taken to solve the CPA equations analytically. However, one could ask about the microscopic relevance of the Lorentzian distribution $P_{\Gamma}(E)$. From the current point of view, $P_{\Gamma}(E)$ makes sense if we realize: (i) The pair-breaking scattering probability distribution should be modeled by the bell-shaped function around the Fermi energy since the scatterings closer to (further apart from) the Fermi level will be more (less) probable. (ii) Empirically, we know that within the Dynes density of states (DOS)~\cite{Dynes78}, the effect of ``quasiparticle energy smearing" interferes with states further apart from the Fermi level. Assuming the energy gap $\Delta$ and the regime $\Gamma\ll\Delta$, it introduces in-gap DOS states\footnote{Already in full width of the gap.} and it also diminishes the coherent peaks located at energy $|\omega| \approx \Delta$. However, the scale of the coherent peaks is defined dominantly by $\Delta$ and not by the characteristic scale $\Gamma$ of the scattering probability distribution. The choice of Lorentzian $P_{\Gamma}(E)$ with diverging root mean square \footnote{Due to slowly decreasing distribution $\propto 1/\omega^{2}$, assuming $\omega\rightarrow\pm\infty$.} fulfills the above criteria. The resulting solution of the CPA equations reproduces the phenomenological Dynes DOS and explains its experimentally motivated artificial parameter $\Gamma=1/(2\tau_{m})\propto n_m$ as pair-breaking scattering rate. Symbol $\tau_{m}$ corresponds to the pair-breaking scattering time and $n_m$ is the particle density of the magnetic impurities \cite{Herman16, Ambegaokar65}.

The simple analytical formulation of the resulting one-particle Green's function~\cite{Herman17_2} creates a convenient basis to address signatures in the more complex experiments. However, so far, a few of the very basic properties of the Dynes superconductor theory were calculated only numerically. 
To fill the gap in the literature, we show that the energy gap behavior (fundamental property of the theory) close to critical temperature $\Delta(T\rightarrow T_c)$ can be achieved analytically. We derive the solution for the general pair-breaking scattering rate with the surprisingly simple result in the critical limit of $\Gamma$. Using our results, the hardware-dependent numerics can be avoided for general pair-breaking and $T$ close to $T_c$.

To clarify, we will mark $T_c$ ($T_{c,0}$) as a critical temperature of the Dynes (BCS, $\Gamma=0$) superconductor. Likewise, we will use $\Delta$ ($\Delta_c$) as the energy gap of the Dynes (BCS - clean) superconductor. In our notation $\Delta(0)~=~\Delta_0$ and $\Delta_c(0)~=~\Delta_{00}$.

The article is structured in the following way. In Sect.~\ref{sec:beh}, we derive the Dynes critical ratio and compare its value to the well-known BCS theory case. In Sect.~\ref{subsec:num_and_an}, we analyze the pair-breaking scattering dependence of the prefactor for $ \Delta(T)/\Delta(0) \propto \sqrt{1 - T/T_c}$. In Sect.~\ref{sec:simulation}, we calculate $\Delta(T)$  numerically and compare it to our analytical results obtained in the Sect.~\ref{subsec:num_and_an}. Section~\ref{sec:conclusions} is dedicated to conclusions.
Detailed calculations can be found in Appendices~\ref{app:for_delta0_delta_00} - \ref{app:Analyt}.

\section{Gap at zero and critical temperature}
\label{sec:beh}
\subsection{Theory}
\label{subsec:theory}
In this section, we focus directly on the Dynes superconductor energy gap $\Delta(T)$ equation formulated in Ref.~\cite{Herman16}. This equation results from the non-zero part of the off-diagonal Green function component containing the superconducting part of the self-energy. It can be also derived together with the self-consistent CPA equations by minimizing corresponding Free energy functional \cite{Herman18}
\begin{equation}\label{eq:delta=2g_pi_t_sum_cez_omega}
\Delta(T) = 2g \pi T  \sum_{\omega_n > 0}^{\Omega} \frac{\Delta(T)}{\sqrt{(\omega_n + \Gamma)^2 + \Delta(T)^2}},
\end{equation}
where $g$ stands for electron-phonon coupling constant, $\omega_n = (2n+1)\pi T$ are Matsubara frequencies and $\Omega$ is the cut-off frequency. To compare the effect of the pair-breaking disorder in our considerations with the BCS theory, we will need the well-known BCS theory ratio \cite{Combescot22} 
\begin{equation}\label{eq:BCS-ratio}
    \Delta_{00}/T_{c,0} = \pi/\gamma_e \approx 1.764,
\end{equation}
where $\gamma_e = e^{\gamma} \approx 1.781$ and $\gamma \approx 0.577$ is the Euler Gamma constant. Solving Eq.~\eqref{eq:delta=2g_pi_t_sum_cez_omega} at $T=\SI{0}{K}$ results in a interesting property: the gap $\Delta_0$ closes at the finite value of pair-breaking scattering as~\cite{Herman16}
\begin{equation}\label{eq:delta_delta00}
    \Delta_0 = \Delta_{00} \sqrt{1 -  2\Gamma/\Delta_{00}}.
\end{equation}
Although this relation is known, for clarity and comprehensiveness, we derive it in Appendix~\ref{app:for_delta0_delta_00}. Notice that the gap closes at the critical limit of $\Gamma = \Delta_{00}/2$.
\subsection{Dynes gap-to-critical temperature ratio \\at critical pair-breaking scattering}
\label{subsec:crit_ratio}
The solution of Eq.~\eqref{eq:delta=2g_pi_t_sum_cez_omega}, provided for clarity step by step in Appendix~\ref{app:digamy}, leads to the known relation \cite{Tinkham} between $T_c$ and $T_{c,0}$ with respect to $\Gamma$ in the form
\begin{equation}\label{eq:digamy}
    \ln\left( \frac{T_c}{T_{c,0}}\right) =  \psi\left(\frac{1}{2}\right) -\psi\left(\frac{1}{2} + \alpha\right),
\end{equation}
where $\psi(z) = -\sum_{k=0}^\infty 1/(z + k)$ is the digamma function and $\alpha = \Gamma/(2\pi T_c)$.\\

{\it Critical limit.} To achieve the analog of the BCS ratio in the Dynes superconductor theory for the critical limit of $\Gamma/\Delta_{00} \rightarrow 1/2$, we need to solve Eq.~\eqref{eq:digamy} for $T_c \rightarrow 0$ and subsequently $\alpha~\rightarrow~\infty$. Let us use $\psi(1/2) = -\gamma - \ln4$ and the Taylor expansion of $\psi(1/2 + \alpha)$ in $\alpha$ up to the second non-zero term: $\psi(1/2 + \alpha) \approx \ln\alpha + 1/(24\alpha^2)$. We are left with
\begin{equation}\label{eq:ln_ln}
    \ln \left( \frac{\Delta_{00}}{2 \Gamma}\right) = \frac{\pi^2}{6} \left( \frac{T_c}{\Gamma}\right)^2,
\end{equation}
where we used Eq.~\eqref{eq:BCS-ratio} to scale our result to the BCS units. In Eq.~\eqref{eq:ln_ln}, $T_c$ is small, so we can use $\Gamma = \Delta_{00}/2$ directly on the right-hand side. After using Taylor expansion of the left-hand side of Eq.~\eqref{eq:ln_ln} in $(1 - 2\Gamma/\Delta_{00}) \ll 1$ up to the first order and some algebra, we are left with the following
\begin{equation}\label{eq:Tc_to_Tc0}
\frac{T_c}{T_{c,0}} = \sqrt{ \frac{3}{2 \gamma_e^2} \left( 1 - \frac{2\Gamma}{\Delta_{00}} \right) },
\end{equation}
which shows the explicit dependence of $T_c$ on $\Gamma$ close to its critical value in a straightforward fashion. Notice that the analog of this relation was assumed in Ref.~\cite{Herman23}, with no clarification whatsoever. Therefore we consider the provided explanation to be a valuable piece of missing reasoning. 
Actually, we can exploit Eq.~\eqref{eq:Tc_to_Tc0} even more by combining it with Eq.~\eqref{eq:delta_delta00}, what leads us directly to the Dynes critical ratio
\begin{equation}\label{eq:Dynes_ratio}
    \Delta_{0}/T_c = \sqrt{2/3}\pi \approx 2.565.
\end{equation}
In comparison with the Eq.~\eqref{eq:BCS-ratio} (clean BCS system), we see that with $\Gamma$ increasing towards its critical value, $T_c$ is suppressed to zero a bit more than the energy gap $\Delta_0$. 

\section{Gap close to critical temperature}
\label{Sec3}

\subsection{Numerical and analytical approaches}
\label{subsec:num_and_an}
In the previous section, we obtained the ratio of the two important energy scales (measured independently) in the limit of pair-breaking being close to its critical value. In addition, it is interesting to find the dependence $\Delta(T)$ close to $T_c$. In the following subsection, we compare the solutions of numerical and analytical approaches, to find out the range of temperatures, where the analytical approximation for $T\rightarrow T_c$ is valid. Therefore, we provide step by step recipe for the numerical calculation of $\Delta(T)$ valid for general $\Gamma$ and $T$ utilizing Eq.~\eqref{eq:delta=2g_pi_t_sum_cez_omega} in Appendix~\ref{app:for_deltaT/delta0}. The resulting equation reads
\begin{align}\label{eq:numerika_pre_delta}
\ln{\frac{T}{T_{c,0}}} =
    \sum_{n=0}^\infty & \frac{T/T_{c,0}}{\sqrt{\left(\left( n + \frac{1}{2}\right) \frac{T}{T_{c,0}} + \frac{1}{2 \gamma_e} \frac{\Gamma}{\Delta_{00}}\right)^2 + \left(\frac{1}{2 \gamma_e} \frac{\Delta(T)}{\Delta_{00}}\right)^2}} \nonumber\\
    &- \frac{1}{n + \frac{1}{2}}.
\end{align}

Second, in Appendix~\ref{app:Analyt}, we derive another main result of this study - the analytical solution for $\Delta(T \rightarrow T_c)$
\begin{equation}\label{eq:deltat_delta0}
    \frac{\Delta(T)}{\Delta_0} =  \frac{\gamma_e T_c}{T_{c,0}} \sqrt{\frac{8(1 - \alpha \zeta(2, \frac{1}{2} + \alpha))}{\zeta(3, \frac{1}{2}+ \alpha) (1 - 2\Gamma/\Delta_{00})} \left(1 - \frac{T}{T_c}\right)}.
\end{equation}
The Eq.~\eqref{eq:deltat_delta0} is true for arbitrary $\Gamma  \leq \Delta_{00}/2 $. Notice the characteristic sign of the mean-field approach provided by the $\sqrt{1-T/T_c}$ dependence. Our main interests are clean and critical $\Gamma$ limits. For the clean limit ($\Gamma = 0$), we achieve the behavior known from the BCS theory\cite{Combescot22}
\begin{equation}\label{eq:delta_Tdelta_00}
    \frac{\Delta(T)}{\Delta_{00}} = \gamma_e \sqrt{\frac{8}{7\zeta(3)}\left(  1 - \frac{T}{T_c}\right)}.
\end{equation}
On the other hand, the critical limit of Eq.~\eqref{eq:deltat_delta0}, when $\Gamma~\rightarrow ~\Delta_{00}/2$ and subsequently $T_c~\rightarrow~0$, meaning $\alpha~\rightarrow~\infty$, completes the information about Dynes superconductor theory. Taylor expansions in $\alpha \rightarrow \infty$ up to the second order of $\alpha$ are the following
\begin{equation*}
    \alpha \zeta(2, 1/2 + \alpha) \approx 1 -  1/(12 \alpha^2),\quad \zeta(3, 1/2 + \alpha) \approx 1/(2\alpha^2).
\end{equation*}
Combining these equations with Eq.~\eqref{eq:deltat_delta0} and Eq.~\eqref{eq:Tc_to_Tc0}, we obtain
\begin{equation}\label{eq:delta_0}
\frac{\Delta(T)}{\Delta_0} = \sqrt{2 \left( 1 - \frac{T}{T_c}\right)}.
\end{equation}

\subsection{Comparison of numerics and analytics}
\label{sec:simulation}
\begin{figure}[b!]
\includegraphics[width=.8\linewidth]{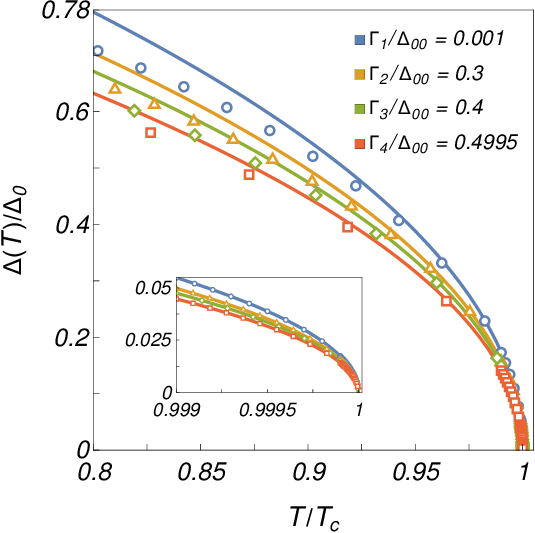}
\caption{Numerical solution of Eq.~\eqref{eq:numerika_pre_delta} (discrete markers) and analytical approximation represented by Eq.~\eqref{eq:deltat_delta0} (solid lines).}
\label{fig:inset}       
\end{figure}
\begin{figure}[h!]
\centering
\includegraphics[width=.8\linewidth]{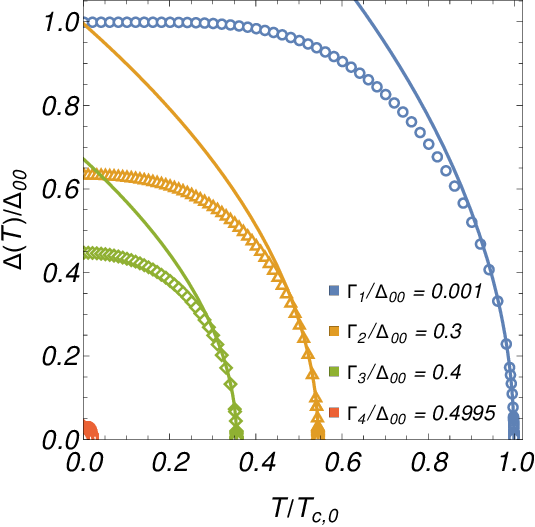}
\caption{Numerical solution of Eq.~\eqref{eq:numerika_pre_delta} (discrete markers) and theoretical approximation Eq.~\eqref{eq:deltat_delta0} (solid line) in ``clean-BCS" units.}
\label{fig:delta_app_ciste}
\end{figure}
\begin{figure}[h!]
\centering
\includegraphics[width=.85\linewidth]{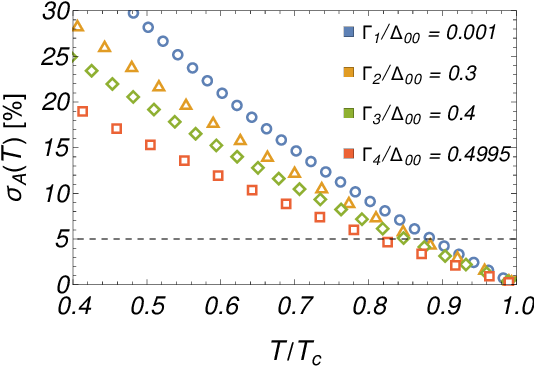}
\caption{The error of the approximative formula Eq.~\eqref{eq:deltat_delta0} for several $\Gamma$ parameters.}
\label{fig:error_app}
\end{figure}

Here, we compare the numerical solution of Eq.~\eqref{eq:numerika_pre_delta}  with analytical dependence provided by Eq.~\eqref{eq:deltat_delta0} for several values of $ \Gamma/\Delta_{00}$. 
In Fig.~\ref{fig:inset}, we see that our approximation corresponds to the numerical solution in the vicinity of $T_c$. In Fig.~\ref{fig:delta_app_ciste}, we can see the dependence $\Delta(T)$ in clean units $(\Delta_{00}, T_{c,0})$. Here the agreement is more clearly visible for each $\Gamma$. In Fig.~\ref{fig:error_app}, we estimate the range of temperatures, where the theory is valid, calculating the relative deviation
\begin{equation}\label{eq:error_formula}
    \sigma_A(T) = |\Delta_A(T)/\Delta_{num}(T)-1|,
\end{equation}
where $\Delta_A(T)$ is our approximative formula from Eq.~\eqref{eq:deltat_delta0} and $\Delta_{num}(T)$ is the numerical solution of Eq.~\eqref{eq:numerika_pre_delta}. Our analytical solution has less than $5\%$ deviation for $T\gtrsim~0.88~T_c$, meaning that our approximative formula applies in this range of temperatures. Notice that our analytical results of $\Delta(T\rightarrow T_c)$ complement the Sommerfeld low-temperature expansion of $\Delta(T)$ published in Ref.~\cite{Herman18}. 


\begin{figure}[t]
\centering
\includegraphics[width=.8\linewidth]{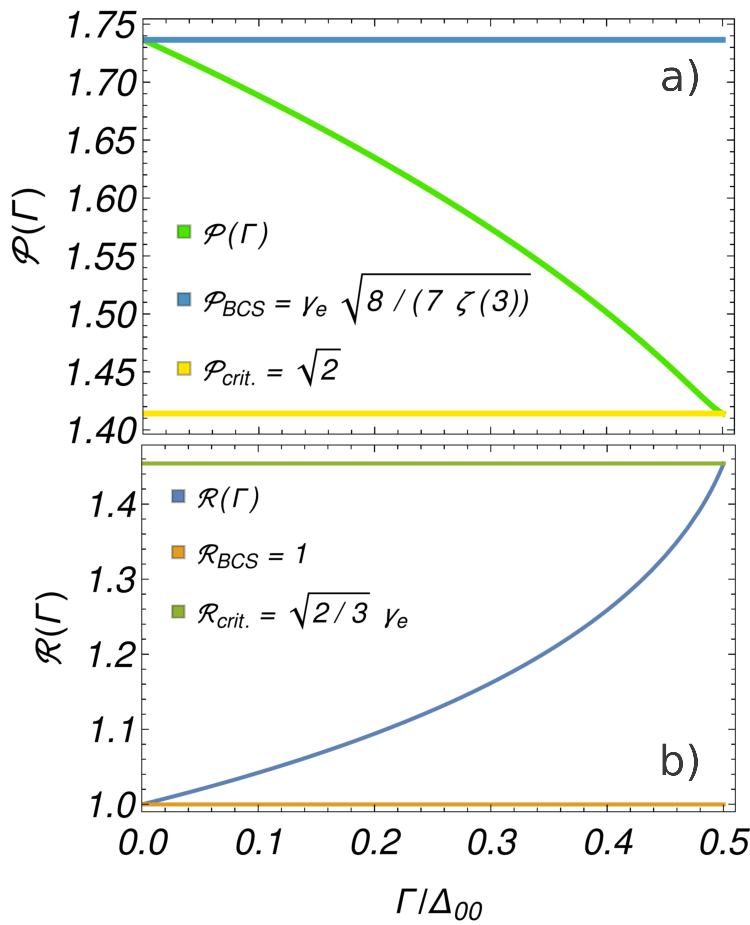}
\caption{a) Behavior of the prefactor $\mathcal{P}(\Gamma)$ together with two limit cases. b) Behavior of the prefactor $\mathcal{R}(\Gamma)$ together with two limit cases.}
\label{fig:panel_1}       
\end{figure}

Fig.~\ref{fig:panel_1}$a)$ depicts the behavior of prefactor $\mathcal{P}(\Gamma)$, defined as $\Delta(T)/\Delta_0 = \mathcal{P}(\Gamma)\sqrt{1 - T/T_c}$ from Eq.~\eqref{eq:deltat_delta0}. As we can see, $\mathcal{P}(\Gamma)$ is a monotonous decreasing function with $\Gamma$. Physically it means that the slope in Fig.~\ref{fig:inset} must decrease, ``flatten" with increasing $\Gamma$ (at fixed temperature the $\Delta$ value decreases with increasing $\Gamma$). This result is in agreement with Fig.~\ref{fig:inset}.

The next important point is that we can analytically compare the ratios $\Delta_0/T_c$ and $\Delta_{00}/T_{c,0}$ depending on $\Gamma$. Let us assume \cite{Abrikos} that $\Delta_0/T_c~= \mathcal{R}(\Gamma) \,\, \Delta_{00}/T_{c,0}$. In Fig.~\ref{fig:panel_1}$b)$, we see that the prefactor $\mathcal{R}(\Gamma)$ increases with increasing $\Gamma$, contrary to $\mathcal{P}(\Gamma)$. In other words, if we rewrite the corresponding equation as $\Delta_0/\Delta_{00} = \mathcal{R}(\Gamma) T_c/T_{c,0}$, we notice that the ratio $\Delta_0/\Delta_{00}$ decreases slower than $T_c/T_{c,0}$, meaning that in dependence shown in Fig.~\ref{fig:delta_app_ciste}, the value $T_c$ is more suppressed than the value $\Delta_0$ with increasing $\Gamma$. In Fig.~\ref{fig:panel_1}a), it is shown that our approximations given by Eq.~\eqref{eq:delta_Tdelta_00} for clean and by Eq.~\eqref{eq:delta_0} for critical pair-breaking limit cases bound the general solution. The same principle for BCS and critical $\Gamma$ limit cases holds true for the behavior of $\mathcal{R}(\Gamma)$ in Fig.~\ref{fig:panel_1}$b)$. Limiting values of $\mathcal{P}(\Gamma)$ turn out to be the same as in Abrikosov-Gorkov theory  \cite{Abrikos, Ambegaokar65, Parks69}.
It is worth mentioning that in \cite{Herman16}, the $\mathcal{R}(\Gamma)$ value was calculated numerically in the critical pair-breaking limit $\mathcal{R}_{crit.}~\approx 1.45$, which corresponds to our precise analytical value $\mathcal{R}_{crit.}~= \sqrt{2/3}\gamma_e~\approx 1.45424$. At the same time, this value is less than the Abrikosov-Gorkov theory \cite{Abrikos, Parks69} prediction $\mathcal{R}_{crit.}^{AG}~= \sqrt{2}\gamma_e~\approx 2.51882$.  

\begin{figure}[t]
\centering
\includegraphics[width=.7\linewidth]{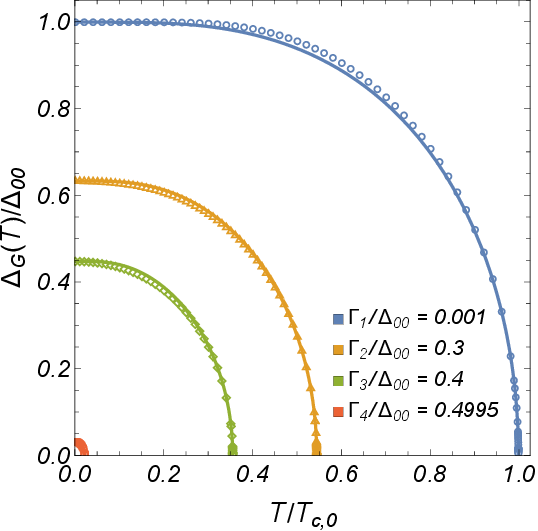}
\caption{Numerical solution of Eq.~\eqref{eq:numerika_pre_delta} (discrete markers) and function $\Delta_{G}(T)$ from Eq.~\eqref{eq:tanh} (solid lines) in the full range of temperature in ``clean-BCS" units.}
\label{fig:tanh_ciste}       
\end{figure}
What is to add, using Eq.~\eqref{eq:deltat_delta0} and the knowledge of the prefactor $\mathcal{P}(\Gamma)$, we heuristically construct the formula, which fits the exact slope for $T \rightarrow T_c$ and fulfills the condition $\lim_{T \rightarrow 0}\Delta(T)/\Delta_0 = 1$ 
\begin{equation}\label{eq:tanh}
\frac{\Delta_G(T)}{\Delta_0} = \tanh \left( {\mathcal{P}(\Gamma) \sqrt{T_c/T - 1}} \right).
\end{equation}

\begin{figure}[h!]
\centering
\includegraphics[width=.8\linewidth]{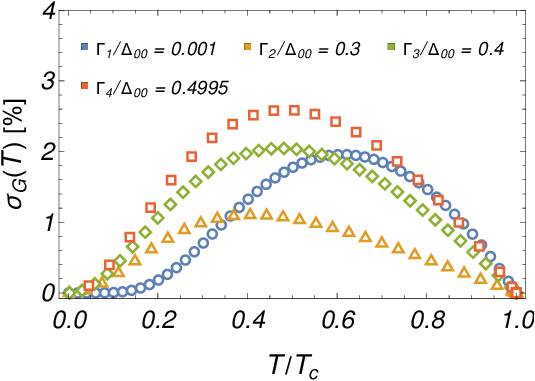}
\caption{The error of the heuristic formula Eq.~\eqref{eq:tanh} for several values of the parameter $\Gamma$.}
\label{fig:error_tanh}
\end{figure}

In Fig.~\ref{fig:tanh_ciste}, we see Eq.~\eqref{eq:tanh} and the previously concerned numerical solution in the full temperature range. 
Although this formula gives the correct value for $T=\SI{0}{K}$, its low-temperature expansion ($\tanh(1/x)\approx 1$ for $x\rightarrow 0$) does not exist in non-zero powers of $T$. This property does not correspond to the $\Delta(T)$ low-temperature expansion found in Ref.~\cite{Herman18}. Therefore, the agreement in the regime $0< T \ll T_c$ is only qualitative. Nevertheless, we examine the relative deviation of our heuristic relation in Fig.~\ref{fig:error_tanh}. We use Eq.~\eqref{eq:error_formula} with $\Delta_G(T)$ instead of $\Delta_A(T)$, resulting in $\sigma_G(T)$. Assuming the full temperature range, the deviation never exceeds $3\%$ boundary. Moreover, for practical values of $\Gamma$, $\sigma_G\leq2\%$ which we find exciting. 

\section{Conclusions}
\label{sec:conclusions}
Exploiting the solution of the gap equation, we achieved the ratio $T_c/T_{c,0}$ in the limit of critical pair-breaking scattering. It led us to the relation of two important superconducting energy scales - Dynes critical ratio, shown in Eq.~\eqref{eq:Dynes_ratio}. Next, we calculated the temperature dependence of the $\Delta(T)$ for $T$ close to $T_c$ for the general case of pair-breaking scattering in Eq.~\eqref{eq:deltat_delta0}, obtaining a surprisingly simple form in the critical limit of $\Gamma$ in Eq.~\eqref{eq:delta_0}. 
In Fig.~\ref{fig:panel_1}, we see that the analytical considerations about the gap behavior made in clean and critical cases agree with the general behavior of the two depicted prefactors $\mathcal{P}(\Gamma)$ and $\mathcal{R}(\Gamma)$. Our analytical results also allow us to easily compare the values of the prefactors in BCS and Dynes critical cases and show the range of applicability of Dynes superconductor theory.
The agreement of the numerical solution of the gap equation and discussed approximative formula is depicted in Fig.~\ref{fig:inset} and Fig.~\ref{fig:delta_app_ciste}. This formula is applicable for $T\gtrsim 0.88 T_c$.
 Based on our exact solutions of the gap equation for $T\approx T_c$, we find the overall dependence $\Delta(T)$, formulated by $\Delta_G(T)$ in Eq.~\eqref{eq:tanh} and shown in Fig.~\ref{fig:tanh_ciste}, providing a simple tool capturing the main effect of pair-breaking disorder in $\Delta(T)$. The deviation analysis shows that this heuristic relation can be generally used instead of cumbersome numerics with the $3\%$ precision.

To sum up, the simple yet reasonable Dynes superconductor theory is shown to be applicable in different situations \cite{Mainwald20, Sindler22, Szabo16, Bagwe24, Simmendinger16}. In this bolt-tightening article, we accomplish the analytical results making the Dynes superconductor theory more concise and potentially much easier to understand and use in theory or experiment.

\begin{acknowledgements}
This work was supported by the Slovak Research and Development Agency under the Contract no. APVV-23-0515 and by the European Union’s Horizon 2020 research and innovation programme under the Marie Skłodowska-Curie Grant Agreement No.~945478.
\end{acknowledgements}




\appendix
\section{Critical value of pair-breaking scattering}\label{app:for_delta0_delta_00}
To comprehend the calculations in Sec.~\ref{sec:beh}, let us exercise derivation of this relation.
We compare the Eq.~\eqref{eq:delta=2g_pi_t_sum_cez_omega} valid for Dynes superconductor with its form in the clean BCS limit, where $\Gamma = 0$
\begin{equation}\label{eq:clean_sum}
    \Delta_c(T) = 2g \pi T  \sum_{\omega_n > 0}^{\Omega} \frac{\Delta_c(T)}{\sqrt{\omega_n^2 + \Delta_c(T)^2}},
\end{equation}
noting that energy gaps in clean and Dynes limits are not the same.
Comparing inverse values of constant $g$ obtained from Eq.~\eqref{eq:delta=2g_pi_t_sum_cez_omega} and Eq.~\eqref{eq:clean_sum}, we are left with
\begin{equation}\label{eq:sums_to_Rieman}
    \sum_{\omega>0}^{\Omega} \frac{2 \pi T}{\sqrt{(\omega_n + \Gamma)^2 + \Delta(T)^2}} = \sum_{\omega>0}^{\Omega} \frac{2\pi T}{\sqrt{\omega_n^2 + \Delta_c(T)^2}}.
\end{equation}
If we take $2\pi T$ to be the integration step, we see it is small in the case of $T \rightarrow 0$, so $\omega_n$ can be understood as a continuous value. Now, we can switch from summation to integration
\begin{equation*}
    \int_0^{\Omega} \frac{d\omega}{\sqrt{(\omega + \Gamma)^2 + \Delta(0)^2}} - \int_0^{\Omega} \frac{d\omega}{\sqrt{\omega^2 + \Delta_c(0)^2}} = 0.
\end{equation*}
Noting that $\Delta(0)~=~\Delta_0$, $\Delta_c(0)~=~\Delta_{00}$ and assuming the cut-off to be high enough $\Omega \rightarrow \infty $, we obtain the solution
\begin{equation*}
    \ln \Biggr[ \left( \Gamma + \sqrt{\Gamma^2 + \Delta_0^2}\right)/\Delta_{00} \Biggl] = 0.
\end{equation*}
From here, we have directly Eq.~\eqref{eq:delta_delta00}.

\section{Gap equation at critical temperature - necessary details}\label{app:digamy}
Let us solve Eq.~\eqref{eq:delta=2g_pi_t_sum_cez_omega} comparing values $1/g$ in clean ($\Gamma = 0$) and Dynes Eq.~\eqref{eq:delta=2g_pi_t_sum_cez_omega} forms at critical temperatures, assuming $N_c = \Omega/(2\pi T_c)$ and $N_{c,0} = \Omega/(2\pi T_{c,0})$
\begin{equation*}
    \frac{1}{g} = \sum_{n=0}^{N_c} \frac{1}{n + \frac{1}{2} + \frac{\Gamma}{2 \pi T_c}} = \sum_{n=0}^{N_{c,0}} \frac{1}{n + \frac{1}{2}}.
\end{equation*}
Noting that $T_c < T_{c,0}$ and assuming $\alpha = \Gamma/(2 \pi T_c)$, one can show that
\begin{equation*}
\sum_{n=N_{c,0}}^{N_c} \frac{1}{n + 1/2 + \alpha} = 
    \sum_{n=0}^{N_{c,0}} \frac{1}{n + 1/2} - \sum_{n=0}^{N_{c,0}} \frac{1}{n + 1/2 + \alpha} .
\end{equation*}
Assuming cut-off scale $\Omega \rightarrow \infty$, we rewrite the sums from $n = 0$ with the help of digamma function  $\psi(z)~=~-\sum_{k=0}^\infty 1/(z + k)$. Assuming high $\Omega$, we change the summation in $n$ to integration on the left-hand side, resulting in 
Eq.~\eqref{eq:digamy}.
\section{Gap equation - formulation for numerics}
\label{app:for_deltaT/delta0}
We begin with Eq.~\eqref{eq:delta=2g_pi_t_sum_cez_omega} for $\Gamma \neq 0$ and switch from summation over $\omega_n$ to $n$, assuming $N = \Omega/(2 \pi T)$
\begin{equation*}
\frac{1}{g} = 2 \pi T \sum_{n = 0}^{N} \frac{1}{\sqrt{(2\pi T (n + \frac{1}{2}) + \Gamma)^2 + \Delta^2}}.
\end{equation*}
Let us compare the right-hand sides of this equation for clean critical temperature and general cases
\begin{equation*}
    \sum_{n=0}^{N_{c,0}} \frac{1}{n + \frac{1}{2}} = \sum_{n=0}^{N} \frac{1}{\sqrt{\left(n + \frac{1}{2} + \frac{\Gamma}{2\pi T}\right)^2 + \left(\frac{\Delta}{2 \pi T}\right)^2}}.
\end{equation*}
Noting that $T < T_c < T_{c,0}$, we divide the sum to the right in two sums and continue with the following
\begin{align*}
    \sum_{n = 0}^{N_{c,0}} \frac{1}{n + \frac{1}{2}} &- \frac{1}{\sqrt{\left(n + \frac{1}{2} + \frac{\Gamma}{2 \pi T}\right)^2 + \left(\frac{\Delta}{2 \pi T}\right)^2}} = \\ 
    &\sum_{n=N_{c,0}}^{N} \frac{1}{\sqrt{\left(n + \frac{1}{2} + \frac{\Gamma}{2 \pi T}\right)^2 + \left(\frac{\Delta}{2 \pi T}\right)^2}}.
\end{align*}
Assuming $\Omega \rightarrow \infty$, we switch from summation to integration on the right-hand side. After simple algebra and using Eq.~\eqref{eq:BCS-ratio} and Eq.~\eqref{eq:Dynes_ratio}, we obtain Eq.~\eqref{eq:numerika_pre_delta}, which can be calculated numerically. It is worth mentioning that the equation above is valid for arbitrary $T < T_c$.

\section{Gap equation - analytical solution close to critical temperature}
\label{app:Analyt}
Now, let us solve the gap equation Eq.~\eqref{eq:delta=2g_pi_t_sum_cez_omega} analytically for $T$ close to $T_c$.
We rewrite the value $1/g$ obtained in general case
\begin{align*}
    &\frac{1}{g} = \sum_{n=0}^{N} \frac{1}{\sqrt{\left(n + \frac{1}{2} + \frac{\Gamma}{2\pi T}\right)^2 + \left(\frac{\Delta}{2 \pi T}\right)^2}} =\\
    &\sum_{n=0}^{N} \frac{1}{\left( n + \frac{1}{2} + \frac{\Gamma}{2 \pi T}\right) \sqrt{1 + \left( \frac{\Delta}{2 \pi T} \frac{1}{n + 1/2 + \Gamma/2 \pi T} \right)^2}}.
\end{align*}
Using Taylor expansion in $ \frac{\Delta}{2\pi T}\frac{1}{n + 1/2 + \Gamma/(2 \pi T)}  \ll 1$ up to the second order, we are left with the following
\begin{equation*}
\frac{1}{g} = \sum_{n = 0}^{N}  \frac{1}{n + \frac{1}{2} + \frac{\Gamma}{2 \pi T}} - \frac{1}{2} \left(\frac{\Delta}{2\pi T}\right)^2 \sum_{n = 0}^{N}  \frac{1}{\left(n + \frac{1}{2} + \frac{\Gamma}{2 \pi T}\right)^3}.	
\end{equation*}
Particularly, at $T = T_c$, we have $\Delta(T_c) = 0$ and obtain
\begin{equation*}
\frac{1}{g} = \sum_{n = 0}^{N_c}  \frac{1}{n + \frac{1}{2} + \alpha}.
\end{equation*}
Once again, we compare the right-hand sides of two previous equations
\begin{multline}\label{eq:tri_sumy}
    \sum_{n = 0}^{N_c}  \frac{1}{n + \frac{1}{2} +\alpha} = \sum_{n = 0}^{N}  \frac{1}{n + \frac{1}{2} + \frac{\Gamma}{2 \pi T}} \\
    - \frac{1}{2} \left(\frac{\Delta}{2\pi T}\right)^2 \sum_{n = 0}^{N}  \frac{1}{\left(n + \frac{1}{2} + \frac{\Gamma}{2 \pi T}\right)^3}.
\end{multline}
Let us look at the first sum on the right-hand side
\begin{equation*}
\sum_{n = 0}^{N}  \frac{1}{n + \frac{1}{2} + \frac{\Gamma}{2 \pi T}} = \sum_{n = 0}^{N}  \frac{1}{n + \frac{1}{2} + \frac{\Gamma}{2 \pi T_c (1 + T/T_c - 1)}}.
\end{equation*}
Assuming $T \rightarrow T_c^-$ and therefore using the Taylor expansion in $\left( 1 - \frac{T}{T_c} \right) \rightarrow 0$ up to the first non-zero term, one can show that
\begin{align*}
\sum_{n = 0}^{N}  \frac{1}{n + \frac{1}{2} + \frac{\Gamma}{2 \pi T}} &= \sum_{n = 0}^{N}  \frac{1}{n + \frac{1}{2} + \alpha + \alpha \left(1 - \frac{T}{T_c} \right)}, \\
&= \sum_{n = 0}^{N}  \frac{1}{n + \frac{1}{2} + \alpha}\left( 1 - \frac{\alpha \left( 1 - T/T_c\right)}{n + 1/2 + \alpha}\right), \\
&=\sum_{n = 0}^{N_c}  \frac{1}{n + \frac{1}{2} + \alpha} + \sum_{n=N_c}^{N}  \frac{1}{n + \frac{1}{2} + \alpha} \\
&- \sum_{n = 0}^{N} \frac{\alpha \left( 1 - \frac{T}{T_c}\right)}{(n + \frac{1}{2} + \alpha)^2}.
\end{align*}
Using this in Eq.~\eqref{eq:tri_sumy} and subtracting equal sums on both sides of the equation, we are left with the following
\begin{multline}\label{eq:pred_Zeta}
    \sum_{n=N_c}^{N}  \frac{1}{n + 1/2 + \alpha} = \alpha \left( 1 - \frac{T}{T_c}\right)\sum_{n = 0}^{N}  \frac{1}{(n + 1/2 + \alpha)^2} \\ 
     +\frac{1}{2} \left(\frac{\Delta}{2\pi T}\right)^2 \sum_{n = 0}^{N}  \frac{1}{\left(n + \frac{1}{2} + \frac{\Gamma}{2 \pi T}\right)^3}.
\end{multline}

Let us now assume the cut-off energy $\Omega$ to be high ($\Omega \gg 1$), so we can use integration instead of summation on the left-hand side
\begin{equation*}
    \sum_{n=N_c}^{N}  \frac{1}{n + 1/2 + \alpha} \approx \int_{N_c}^{N} \frac{1}{n + 1/2 + \alpha} dn = \ln \left( \frac{T_c}{T}\right).
\end{equation*}
Now looking at the right-hand side of Eq.~\eqref{eq:pred_Zeta} and using Zeta-function $\zeta(a, b) = \sum_0^\infty 1/(n+b)^a$, we obtain
\begin{align*}
    \ln \left( \frac{T_c}{T}\right) &= \alpha \zeta\left(2, \frac{1}{2} + \alpha\right) \left( 1 - \frac{T}{T_c}\right)\\ 
    &+ \frac{1}{2} \zeta\left(3, \frac{1}{2} + \alpha \frac{T_c}{T}\right) \left(\frac{\Delta}{2\pi T}\right)^2.
\end{align*}
Adding and subtracting $1$ in the logarithm argument, then using Taylor expansion in $(1 - T/T_c) \ll 1$ up to the first non-zero term and assuming $\Delta \rightarrow 0$, we obtain
\begin{multline*}
    \left( 1 - \alpha \zeta\left(2, \frac{1}{2} + \alpha\right) \right)\left( 1 - \frac{T}{T_c}\right) = \\
     = \frac{1}{2} \left(\frac{\Delta}{2\pi T_c}\right)^2 \zeta\left(3, \frac{1}{2} + \alpha\right).
\end{multline*}
In the natural units of $\Delta_0$, we are left with
\begin{equation*}
\frac{\Delta(T)}{\Delta_0} = \frac{\pi T_c}{\Delta_0} \sqrt{\frac{8(1-\alpha \zeta(2, \frac{1}{2}+ \alpha))}{\zeta(3, \frac{1}{2} + \alpha)} \left(1-\frac{T}{T_c}\right)}.
\end{equation*}
(Re)arranging by the use of Eq.~\eqref{eq:BCS-ratio} and Eq.~\eqref{eq:delta_delta00}, we obtain Eq.~\eqref{eq:deltat_delta0}.


\begin{thebibliography}{}

\bibitem{Herman16}Herman, F. and Hlubina, R.: Microscopic interpretation of the Dynes formula for the tunneling density of states. Phys. Rev. B {\bf 94}, 144508 (2016). https://doi.org/10.1103/PhysRevB.94.144508

\bibitem{Herman17_1}Herman, F. and Hlubina, R.: Consistent two-lifetime model for spectral functions of superconductors. Phys. Rev. B {\bf 95}, 094514 (2017). https://doi.org/10.1103/PhysRevB.95.094514

\bibitem{Herman17_2}Herman, F. and Hlubina, R.: Electromagnetic properties of impure superconductors with pair-breaking processes. Phys. Rev. B {\bf 96}, 014509 (2017). https://doi.org/10.1103/PhysRevB.96.014509

\bibitem{Herman18}Herman, F. and Hlubina, R.: Thermodynamic properties of Dynes superconductors. Phys. Rev. B {\bf 97}, 014517 (2018). https://doi.org/10.1103/PhysRevB.97.014517

\bibitem{Soven67}Soven, P.: Coherent-Potential Model of Substitutional Disordered Alloys. Phys. Rev. {\bf 156}, 809 (1968). https://doi.org/10.1103/PhysRev.156.809

\bibitem{Velicky68}Velický, B., Kirkpatrick, S., Ehrenreich, H.: Single-Site Approximations in the Electronic Theory of Simple Binary Alloys. Phys. Rev. {\bf 175}, 747 (1968). https://doi.org/10.1103/PhysRev.175.747

\bibitem{Weinkauf75}Weinkauf, A., Zittartz, J.: Theory of superconducting alloys. J. Low Temp. Phys. {\bf 18}, 229 (1975).
https://doi.org/10.1007/BF00118155

\bibitem{Szabo16}Szabó, P., Samuely, T., Hašková, V., Kačmarčík, J., Žemlička, M., Grajcar, M., Rodrigo, J.G., Samuely, P.: Fermionic scenario for the destruction of superconductivity in ultrathin MoC films evidenced by STM measurements. Phys. Rev. B {\bf 93}, 014505 (2016). https://doi.org/10.1103/PhysRevB.93.014505

\bibitem{Zemlicka20}Zemlička, M., Kopčík, M., Szabó, P., Samuely, T., Kačmarčík, J., Neilinger, P., Grajcar, M., and Samuely, P.: Zeeman-driven superconductor-insulator transition in strongly disordered MoC films: Scanning tunneling microscopy and transport studies in a transverse magnetic field. Phys. Rev. B {\bf 102}, 180508(R) (2020). https://doi.org/10.1103/PhysRevB.102.180508

\bibitem{Mainwald20}Maiwald, J., Mazin, I.I., Gurevich, A., Aronson, M.: Superconductivity in La2Ni2In. Phys. Rev. B {\bf 102}, 165125 (2020). https://doi.org/10.1103/PhysRevB.102.165125

\bibitem{Sindler22}Šindler, M., Kadlec, F., Kadlec, C.: Onset of a superconductor-insulator transition in an ultrathin NbN film under in-plane magnetic field studied by terahertz spectroscopy. Phys. Rev. B {\bf 105}, 014506 (2022). https://doi.org/10.1103/PhysRevB.105.014506

\bibitem{Herman23}Herman, F. and Hlubina, R.: Slope of $H_{c2}$ close to $T_{c}$ versus the size of the Cooper pairs: The role of disorder in Dynes superconductors. Phys. Rev. B {\bf 108}, 134518 (2023). https://doi.org/10.1103/PhysRevB.108.134518

\bibitem{Dynes78}Dynes, R. C., Narayanamurti, V., Garno, J. P.: Direct Measurement of Quasiparticle-Lifetime Broadening in a Strong-Coupled Superconductor. Phys. Rev. Lett. {\bf 41}, 1509 (1978). https://doi.org/10.1103/PhysRevLett.41.1509

\bibitem{Ambegaokar65}Ambegaokar, V., Griffin, A.: Theory of the Thermal Conductivity of Superconducting Alloys with Paramagnetic Impurities. Phys. Rev. {\bf 137}, 4A (1965).

\bibitem{Combescot22}Combescot, R.: \textit{Superconductivity, An Introduction}, p.~98. Cambridge University Press, Cambridge (2022).

\bibitem{Tinkham}Tinkham, M.: \textit{Introduction to superconductivity}, p.392, Dover, New York (2004).

\bibitem{Abrikos}Abrikosov, A. A., Gorkov, L. P.: Contribution to the Theory of Superconducting Alloys With Paramagnetic Impurities, Sov. Phys. JETP {\bf 12}, 1243 (1961).

\bibitem{Parks69}Parks, R. D.: {\it Superconductivity}, Vol. 2, p. 1044, MARCEL DEKKER, INC., New York (1969) and references therein.


\bibitem{Bagwe24}Bagwe, V., Duhan, R., Chalke, B., Parmar, J., Basistha, S., Raychaudhuri, P.: Origin of superconductivity in disordered tungsten thin films. Phys. Rev. B {\bf 109}, 104519 (2024). https://doi.org/10.1103/PhysRevB.109.104519

\bibitem{Simmendinger16}Simmendinger, J., Pracht, U. S., Daschke, L., Proslier, T., Klug., J. A., Dressel, M., Scheffler, M.: Superconducting energy scales and anomalous dissipative conductivity in thin films of molybdenum nitride. Phys. Rev. B {\bf 94}, 064506 (2016). https://doi.org/10.1103/PhysRevB.94.064506

\end{thebibliography}
\end{document}